\documentclass[11pt,twoside]{article}

\usepackage{asp2010}
\usepackage{epsf}
\usepackage{lscape}
\usepackage{natbib}

\begin{document}
\title{Spectral Analysis of Sunspot Penumbrae Observed with HINODE} 
\author{Morten Franz, Rolf Schlichenmaier} 
\affil{Kiepenheuer Institut f\"ur Sonnenphysik, Sch\"oneckstra$\beta$e. 6, 79104 Freiburg, Germany}

\begin{abstract}
To investigate the penumbral plasma flow on a small scale, spectropolarimetric data of sunspots recorded by HINODE was used. Maps of Doppler velocities were created by evaluating the bisector in the line-wing, thereby visualizing the flow pattern in the low photosphere where the Evershed effect is most pronounced.

In penumbrae close to the disk center, the vertical component of the Evershed flow dominates. The latter consists of a series of elongated up-flow patterns extending radially through the entire center-side penumbra at a constant azimuth. Along this structure, strong up-flows appear in concentrated patches separated by weaker up-flows or even down-flows. The strong up-flows appear at the bright heads and the umbral side of the dark-core of the filament, while the down-flows are rather located at the penumbral side of the filament. Projection effects lead to an overall red-shift of the limb-side penumbra, but the described pattern of up- and down-flows is still ascertainable.
\end{abstract}


\section{Introduction}
Although penumbrae of sunspots have been subject to scientific investigation for a long time, there remain a lot of open questions regarding  e.g. their filamentary fine structure, the energy transport and the plasma dynamics on a small scale. 

The {\it{gappy}} model \citep{Spruit2006} assumes the existence of field-free, radially aligned gaps below the $\tau=1$ level intruding into a potential field above. Dark cores of penumbral filaments \citep{Scharmer2002} are explained as a variation of the $\tau=1$ level caused by strong magnetic fields in the areas enclosing the gap. Even though the {\it{gappy}} model provides an explanation not only for the dark cores but also for the brightness of the penumbra, it poses a problem as it does not explain the Evershed flow. 

In the {\it{uncombed}} model \citep{Solanki1993}, nearly horizontal flux tubes are assumed to be embedded in a less inclined magnetic background field \citep{Schlichenmaier1998a,Schlichenmaier1998b}. The strong gradients in atmospheric parameters are encountered for any line of sight (LOS) penetrating both the background atmosphere and the flux tube. It has recently been shown that dark cores of penumbral filaments can be explained within the framework of this model by taking the hot Evershed flow embedded in a stratified atmosphere into account \citep{RuizCobo2008}. Despite the success of the {\it{uncombed}} model explaining e.g. the Evershed flow, the existence of bright filaments and their migration, it does not offer a self-consistent model for the penumbral energy transport.

It thus seems crucial to investigate the plasma flow in the low photosphere on a small scale in order to understand the convective nature of the energy transport in the inclined magnetic field of the penumbra.

\section{Observation \& Data Analysis}
For the investigation, data obtained by the spectropolarimeter (SP) \citep{Lites2001} of the solar optical telescope (SOT) \citep{Tsuneta2007} onboard the HINODE satellite was used. The SP records the entire stokes spectra of the two iron lines at 630.15 nm and 630.25 nm, their Lande factors being $g=1.67$ and $g=2.5$ respectively. By scanning the spectrograph slit across the target in steps of 0$^{\prime\prime}$.148 and achieving a wavelength sampling of 2.15 pm/pixel, two dimensional maps are obtained. The width of the slit is equivalent to 0$^{\prime\prime}$.16, thus normal SP scans provide a spatial resolution of 0$^{\prime\prime}$.32. For an exposure time of 4.8 s per slit position, the noise level in the Stokes spectra is of the order of $10^{-3}{\cdot}I_c$, with $I_c$ being the continuum intensity. 

\begin{figure}[htb]
\begin{center}
 \plotone{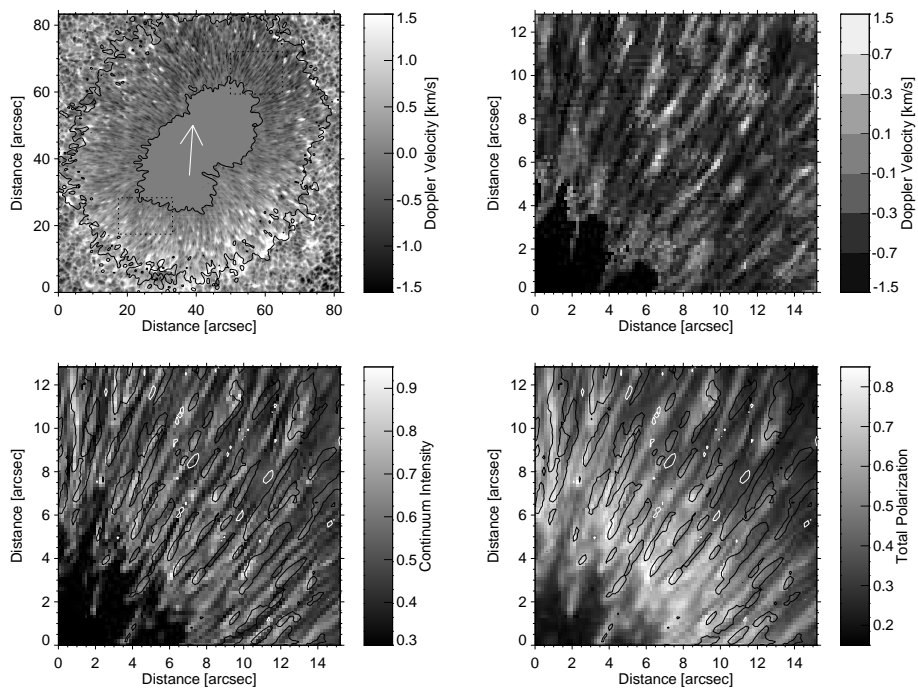}
 \caption{Top: Doppler maps of NOAA 10923 at $\Theta$=$8^{\circ}$. In the picture of the entire spot (left), the arrow points towards disk center, umbra and penumbra were enclosed by black contours, and the dotted boxes mark the area under study. In the detailed picture of the center-side penumbra (right), a range of velocities has been assigned to one gray scale. Bottom: Map of continuum intensity (left) and total polarization (right). The contours mark up-flows $\leq$ -600 m s$^{-1}$ (black) and down-flows $\geq$ 100 m s$^{-1}$ (white).}
 \label{Franz_fig1}
 \end{center}
 \end{figure}

 \begin{figure}[htb]
\begin{center}
 \plotone{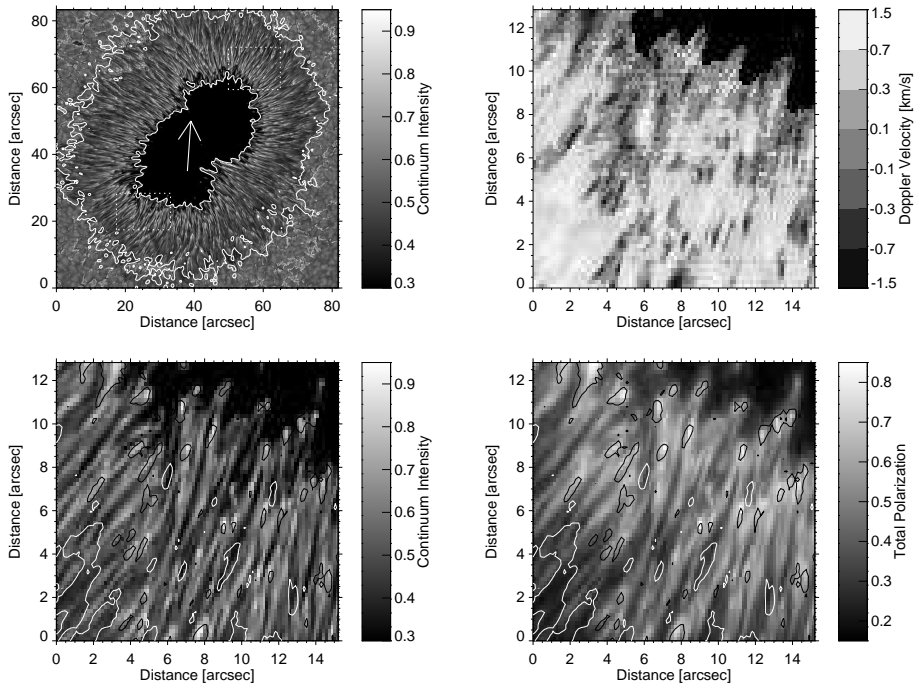}
 \caption{Top: Continuum picture of NOAA 10923 at $\Theta$=$8^{\circ}$ (left) and detailed Doppler map of limb side penumbra (right). Bottom: Map of continuum intensity (left) and total polarization (right). The contours mark up-flows $\leq$ -200 m s$^{-1}$ (black) and down-flows $\geq$ 600 m s$^{-1}$ (white).}
 \label{Franz_fig2}
 \end{center}
 \end{figure}
 
On November 14 2006 at 07h UT, {\it{Hinode}} observed sunspot NOAA 10923 almost at disk center, ${\Theta} = 8^{\circ}$. From the Stokes spectra of Fe 630.15 nm maps of continuum intensity $I_c$, Doppler velocity along the LOS $v_{dop}$ and total polarization$P_{tot}=\int [(Q+U+V)/I_c)^2]^{1/2}d\lambda$ were computed. To create the maps of $v_{dop}$, the bisector of the Fe 630.25 nm line was calculated with sub-pixel accuracy. To calibrate the velocity scale, an average blue-shift of the quiet Sun of -262 m s$^{-1} $ \citep{Borrero2002, Beck2005} was taken into account. Since the bisector was evaluated in the far line-wing, the Doppler images display the plasma velocities in the low photosphere where the Evershed effect is most pronounced.

\section{Results}
In the {\it{center-side penumbra at small heliocentric angles}} an alternating pattern of bright and dark features, which accounts for the filamentary structure of the penumbra, is visible in the continuum intensity. Additionally, the bright heads of some filaments extending into the umbra are observable. This pattern is even more pronounced in the picture showing the total polarization, where the dark cores (areas of low $P_{tot}$) are located in between the lateral brightening (areas of high $P_{tot}$). Since $\Theta$ is only $8^{\circ}$, the maps of Doppler shifts are dominated by plasma flows normal to the surface. The up-flow appears as a series of radially aligned elongated structures. 

Strong up-flows ($\leq$ -600 m s$^{-1}$) are present, separated by weaker up-flows ($>$ -600 m s$^{-1}$) or even down-flows ($>$0 m s$^{-1}$). It also appears that the up-flow is rather located at the bright head and at the umbral side of the dark core of the filament. This up-flow becomes weaker or even turns into a down-flow as one looks along the dark core towards the quiet Sun. This can be seen e.g. in the filament located at $x=4^{\prime\prime}-7^{\prime\prime}$ \& $y=2^{\prime\prime}-7^{\prime\prime}$ in Fig. \ref {Franz_fig1}. 

In the {\it{limb-side penumbra at small heliocentric angles}}, the alternating pattern of bright and dark features is seen in the images of continuum intensity and total polarization respectively. The flow pattern inside a single dark core of a filament is comparable to that in the center-side penumbra, even though it is obscured by the horizontal Evershed outflow, leading to an overall red-shift. Again, the up-flow appears at the bright head of the filament and the down-flow, which appears stronger, is rather at the end of the dark core (cf. Fig. \ref{Franz_fig2}). In the outer penumbra, a strong down-flow is predominant, but it is more dislocated and spread out over a wide area.

\section{Conclusion}
The vertical flow pattern in sunspot penumbrae is structured on a small scale. The appearance of down-flows not only in the outer penumbra, but at all radial distances especially in the inner and mid penumbra is remarkable. These findings are consistent with the results of other observations \citep{Ichimoto2007}. The radially aligned, elongated up-flow patches are of different flow strengths, a feature that is also seen in simulation \citep{Rempel2008}. Sometimes this up-flow pattern is interrupted by down-flow patches. Additionally, it seems that the up-flow appears rather at the bright head and at the umbral side of the dark core of the filament and that the down-flow coincides with the end of the filament pointing towards the quiet Sun.

\end{document}